\documentstyle[prl,aps,psfig,floats,twocolumn]{revtex}
\begin{document}
\draft
\twocolumn[\hsize\textwidth\columnwidth\hsize\csname@twocolumnfalse\endcsname
\title{Topology of correlation based minimal spanning trees in real and model markets}
\author{Giovanni Bonanno$^{1,2}$, 
Guido Caldarelli$^{1}$, Fabrizio Lillo$^{3}$ \and 
Rosario N. Mantegna$^{2,3}$} 
\address{ $^{1}$ Unit\`a INFM di Roma1, Dipartimento di Fisica Universit\`a "La Sapienza" \\
Piazzale A.Moro 5, I-00185 Roma, Italy.\\
$^{2}$ Dipartimento di Fisica e Tecnologie Relative, Universit\`a di Palermo,
viale delle Scienze, I-90128 Palermo, Italy\\
$^{3}$ Unit\`a INFM di Palermo, viale delle Scienze, I-90128 Palermo, Italy \\}
\date{\today} 
\maketitle 

\begin{abstract} 
We present here a topological characterization of the 
minimal spanning tree that can be
obtained by considering the price return correlations of 
stocks traded in a financial market.
We compare the minimal spanning tree obtained from a large group of stocks traded 
at the New York Stock Exchange during a 12-year trading period
with the one obtained from surrogated data simulated by using simple market models. 
We find that the empirical tree has features of a complex network that cannot 
be reproduced, even as a first approximation, by a random market model and by the 
one-factor model.
\end{abstract}
 ]
]
\narrowtext

The study of topological properties of networks has recently received
a lot of attention. In particular it has been shown that many natural 
systems display an unexpected amount of correlation\cite{watts} 
with respect to traditional models\cite{ER}. 
Graphs are mathematical objects formed by vertices connected by 
arcs. An important characterization of a graph is given 
by the degree of vertices, that is the number of arcs per vertex. 
In an Erd\H{o}s-R\'enyi random graph the degree distribution
has a poissonian form, whereas in many cases of 
interest ranging from the  
WWW\cite{JB} to the Internet\cite{CMP,PV} to social networks\cite{N} 
the degree is power law distributed.
The presence of such scale free behavior has been put in 
relationship with the basic ingredient of the network growth 
in time\cite{BA}. We present here an example of system whose size does not 
increase in time but still displays a scale free behavior 
(see also \cite{GC}). 
The case of study is given by a network 
whose vertices are a fixed number of stocks continuously traded at the 
New York Stock Exchange 
(NYSE) and the arcs are obtained by considering the return
cross-correlations. The network is a Minimal Spanning Tree (MST)
connecting all the stocks.
Spanning trees are particular types of graphs. They connect all the 
vertices in a graph without forming any loop.
Therefore if the number of vertices is $n$, one has 
$n-1$ arcs to connect them. There are several example of spanning 
trees in nature and several 
observables have been proposed in order to classify them
and study the possible optimization with respect to some external 
cost function\cite{Maritan}. 

In this letter we study the topological properties of the 
MST obtained from the return cross 
correlation matrix of a portfolio of financial assets. 
We address the problem of the topological structure of the MST by comparing
the real data with the MST obtained from
simple models of the portfolio dynamics. Specifically we
consider a model of uncorrelated Gaussian return time 
series and the widespread one-factor model.
This last model is the starting point of the Capital
Asset Pricing Model\cite{campbell}.

The topological characterization of the correlation based MST
of real data has been already studied in Ref. \cite{vandewalle}. 
Differently from this approach, here we use a smaller 
number of stocks $N$ but we use a number of time records $T$ larger 
than the number of stocks. Our choice is motivated by the 
request that the correlation matrix be positive definite.
When the number of variables is larger than the number of time records the covariance
matrix is only positive semi-definite\cite{Mardia}. Moreover, the application 
of the random matrix theory to the spectral properties of the 
correlation matrix can be applied only when $T/N>1$.

The variable under investigation is the daily price return $r_i(t)$ 
of asset $i$ on day $t$.
Given a portfolio composed of $N$ assets traded simultaneously in
a time period of $T$ trading days, 
we extract the $N\times N$ correlation matrix.
Each correlation coefficient $\rho_{i,j}$ can be associated to
a metric distance $d_{i,j}=\sqrt{2(1-\rho_{i,j})}$ between asset $i$ and $j$ through the
relation \cite{gower,mantegna99}.
The distance matrix is then used to determine the 
MST connecting all the assets.
The method of constructing the MST linking $N$ objects is 
known in multivariate analysis as the nearest neighbor single 
linkage cluster algorithm\cite{Mardia}. 
In a previous study three of us showed that the structure of the 
MST changes with the time horizon used to compute price returns
\cite{bonanno2001}.

The data set used here consists of daily closure prices for 1071 stocks 
traded at the NYSE and continuously present in the 12-year period 1987-1998 
(3030 trading days). It is worth noting that the ratio $T/N \simeq 2.83$ is
significantly larger than one.
With our choice the correlation matrix is positive definite and the theoretical
results of the random matrix theory are valid.
Figure 1 shows the MST of the real data. 
The color code is chosen by using the main industry sector of each 
firm according to the Standard Industrial Classification 
system\cite{sic} for the main industry sector of each firm and the correspondence is 
reported in the figure caption.
Regions corresponding to different sectors are clearly seen. 
Examples are clusters of stocks belonging to the financial sector (purple), 
to the transportation, communications, electric gas and sanitary services sector (green)
and to the mining sector (red). The mining sector stocks are observed to belong
to two subsectors one containing oil companies (located on the right side of the figure)
and one containing gold companies (left side of the figure).

The empirical MST of real data can be compared with the results 
obtained from simple models of the simultaneous dynamics of
a portfolio of assets. 
The simplest model assumes that the return time 
series are uncorrelated Gaussian time series, i.e. 
$r_i(t)=\epsilon_i(t)$, where $\epsilon_i(t)$ are
Gaussian random variables with zero mean and unit
variance.
This type of model has been considered in 
Ref. \cite{JP,Gopi} as a null hypothesis in the study of
the spectral properties of the correlation
matrix.  In the cited references it has been shown
that the spectrum of the real correlation matrix
has a very large eigenvalue corresponding to the 
collective motion of the assets.
A random model does not explain this empirical observation and therefore
this fact clarifies why a better modeling of the portfolio
dynamics is obtained by using the one-factor model.
The one-factor model assumes that the return of assets 
is controlled by a single factor (or index).
Specifically for any asset $i$ we have
\begin{equation}
r_i(t)=\alpha_i+\beta_i r_M(t)+\epsilon_i(t),
\label{onefact}
\end{equation}
where $r_i(t)$ and $r_M(t)$ are the return of the asset $i$ and of the
market factor at day $t$ respectively, $\alpha_i$ and $\beta_i$ are two
real parameters and $\epsilon_i(t)$ is a zero mean noise term
characterized by a variance equal to $\sigma^2_{\epsilon_i}$.
Our choice for the market factor is the Standard \& Poor's 500 index and we assume
that $\epsilon_i=\sigma_{\epsilon_i}w$, where $w$ is a random variable
distributed according to a Gaussian distribution.

We estimate the model parameters for each asset from real time 
series with ordinary least squares method\cite{campbell} and we
use the estimated parameters to 
generate an artificial market according to 
Eq. (\ref{onefact}).
A consequence of this equation is that the variance 
(the squared volatility) of asset $i$ can be written as 
the sum of a term depending on the market factor and 
an idiosyncratic term.
The fraction of variance explained by the factor $r_M$
is approximately described by an exponential distribution 
with a characteristic scale of about 0.16.
The random model can be considered as the limit of the one factor model when
the fraction of variance explained by the factor goes to zero.

In the MST obtained with the random model few nodes have
a degree larger than few units. This implies that the MST is 
composed by long files of nodes. These files join at nodes
of connectivity equal to few units.
The MST obtained with the one-factor model is very different from
the one obtained with the random model. 
In Figure 2 we show the MST obtained in a typical realization  
of the one-factor model performed with the control
parameters obtained as described above.
It is evident that the structure of sectors of Fig.~1 is not present in Fig.~2. 
In fact the MST of the one-factor model has a star-like structure
with a central node. 
The largest fraction of node links directly to the central node and 
a smaller fraction is composed by the next-nearest neighbors.
Very few nodes are found at a distance of three links from 
the central node.   
The central node corresponds to General Electric and the second
most connected node is Coca Cola. It is worth noting that these two stocks
are the two most highly connected nodes in the real MST also.

The MSTs obtained by simulating the models are different in each realization.
However a statistical characterization of MST is possible.
In order to characterize quantitatively the structure of the MST
we make use of two topological quantities. 
The first one is the distribution of the degree $k$. 
In random graph this quantity is distributed 
according to a binomial distribution which for large
networks tends to a Poisson distribution.
In many real networks it has been shown that the degree
is distributed according to power law distribution
signaling the presence of long range correlation.
The second topological quantity is frequently used for 
oriented graphs. 
For any vertex $i$ in the tree we count the total 
number of vertices $a$ in the uphill subtree whose root is $i$.
This quantity is called drainage basin area in oriented
graphs of river networks\cite{IR}, 
whereas it is usually referred as the in-degree component
in graph theory.
To calculate the in-degree component in a correlation based MST,
we orient the MST 
according to the number of steps each node is far from
the most connected node (sink).
When more than one sink is present in the MST a preferential one
is randomly chosen among them.

We report in Figure 3 the frequency distribution for the degree $k$
for the real data and for the average over $100$ realizations
of the random model and of the one factor model.
The degree distribution for the MST of the real data 
shows a power law behavior with exponent $-2.6$ for one decade followed by 
a set of isolated points with high degree.
A power law behavior with a similar exponent has been observed
in Ref.\cite{vandewalle} and in another recent study\cite{kertesz}.
The highest degree $k_{max}=115$ is observed for the General Electric,
one of the most capitalized company in the NYSE. 
As we pointed out in a previous work\cite{bonanno2001}, some 
important companies clearly emerge for its high 
degree value indicating that they act as a reference for other companies. 
The random model displays an approximately exponential decay
of the degree distribution. The value of the maximum 
degree is small, $k_{max}=7.34\pm0.92$, showing that no asset
plays a central role in the MST.
The correlation based MST of the random model can be considered
as the MST of a set of $N$ points randomly distributed in 
an Euclidean space with $d=T$ dimension \cite{MS}. 
The $N$ points have independent identically  
Gaussian distributed coordinates 
${\bf r_i}=(r_i(1),r_i(2),...,r_i(T))$ with $i=1,2,...,N$.
It has been shown that 
the distribution of degree of the random MST in Euclidean space 
converges to a specific distribution in the mean field limit $d\to\infty$\cite{penrose}. 
The numerical values of the degree frequency
obtained from this mean field limit are shown as a
star in Figure 3 for $k=1,..,7$. The agreement of theoretical values with the numerical simulations 
is very good showing that the mean field 
limit is already a good approximation for our T parameter.   

The MST obtained from the one factor model  
is characterized by a rapidly power-law decaying degree distribution and
by an asset with a very high value of the degree, which is
indicated by an arrow in Fig.~3.
The value of the maximum degree is $k_{max}=718\pm29$.
The corresponding asset is the center of the star like 
structure of Figure 2.
The region with highest value of the degree contains information
about the stocks that act as reference for a large set of other stocks.
To get more insight in the structure of this high $k$ region
we show a rank plot of the degree both for
real data and for the considered models in the inset of Figure 3.
For the real market it is evident the presence of a region of power law 
extending for more than one decade.
On the other hand, for the random model many nodes have a similar
value of the degree which is ranging for less than an order 
of magnitude. This is due to the fact that
there is no hierarchy in the random model. The rank plot of the degree of the MST for the one 
factor model has not a scale free behavior. Indeed, there is a single highly 
connected node (the center) and a rapidly decaying degree as a function
of the rank. This fact corresponds to the simple one-center
hierarchy of the MST of the one-factor model.

A discrepancy between real data and models is also observed in the frequency
distribution of the in-degree component.
Fig. 4 shows the frequency distribution of the in-degree component for
real and surrogate data. The inset of Figure 4 shows the rank plot of
the same data. 
In all three cases the in-degree component distribution  has a power law shape.
This is particularly clear for the MST of the random uncorrelated time
series where the power law last for more than two decades with an
exponent of approximately $-1.6$.
It is known that for critical random trees the probability distribution
of tree size decays as a power-law with an exponent $3/2$ \cite{harris}. A
critical random tree is a tree in which the mean number of sons of each
node is one.
In a MST the mean degree is exactly equal to $2n/(n-1) \simeq 2$.
Hence when we orient
the MST from the root to the leaves we have a tree with one son for each
node. 
Our result shows that the in-degree component of the MST arising
from random uncorrelated time series has properties similar to the
one of a critical random tree.
This is not the case for the one-factor model where the power law has
greater absolute slope due to the star-like structure of the tree.
Neither models is actually able to catch the oriented structure
of real data whose in-degree component distribution is in between the two models.  
The same arguments are also valid for the region of high values of $a$ as is evident
from the rank plot in the inset.

In summary these results show that the topology of the MST for the real and 
for the considered artificial markets is different for node with both high
and low degree. 
If we define the importance of a node as its degree 
(or its in-degree component), from our analysis emerges that the real market 
has a hierarchical distribution of importance of the nodes 
whereas the considered models are not able to catch such a
hierarchical complexity. Specifically, in the random model 
the fluctuations select randomly few nodes and assign them
small values of degree. Thus the MST of the random model is
essentially non hierarchical. On the other hand the MST
of the one factor model shows a simple one-center hierarchy.
The MST of real market shows a more structured hierarchy of
the importance of the stocks which is not captured by the
considered models.
The topology of stock return correlation based MST
shows large scale correlation properties characteristic of 
complex networks in the native as well as in an oriented form. 
Such properties cannot be reproduced at all, even as a first 
approximation, by simple models as a random model or the
widespread one-factor model.

Authors acknowledge partial support of 
FET Open project COSIN IST-2001-33555. 
F.L. and R.N.M. acknowledges partial support from INFM and MIUR.



\begin{figure}
\centerline{\psfig{file=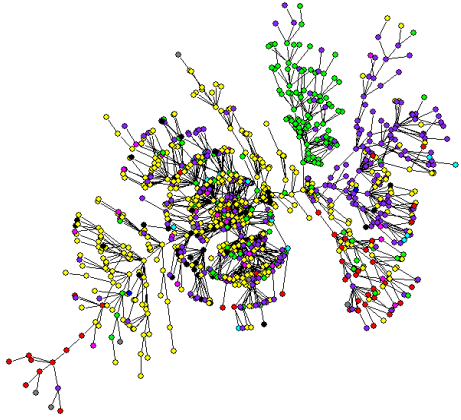,width=6.5cm}}
\caption{Correlation based minimal spanning tree of real data 
from daily stock returns of 1071 stocks for the 12-year period 
1987-1998 (3030 trading days). The node color is
based on Standard Industrial Classification system. The correspondence is: red for mining - cyan for construction - 
yellow for manufacturing - green for transportation, communications, 
electric, gas and sanitary services - magenta for wholesale trade - black for 
retail trade - purple for finance, insurance and real estate - 
orange for service industries - light blue for public administration}
\label{fig1}
\end{figure}

\begin{figure}
\centerline{\psfig{file=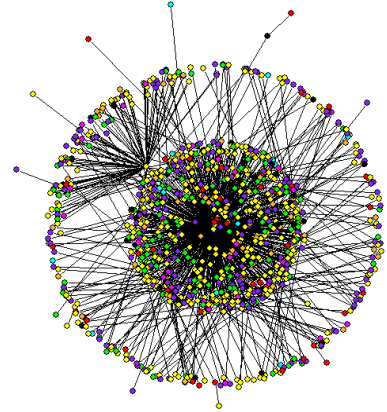,width=6.5cm}}
\caption{Correlation based minimal spanning tree of a numerical simulation
of the one factor model of Eq.(\ref{onefact}).
The color code is the same used in Figure~1}
\label{fig2}
\end{figure}

\begin{figure}
\centerline{\psfig{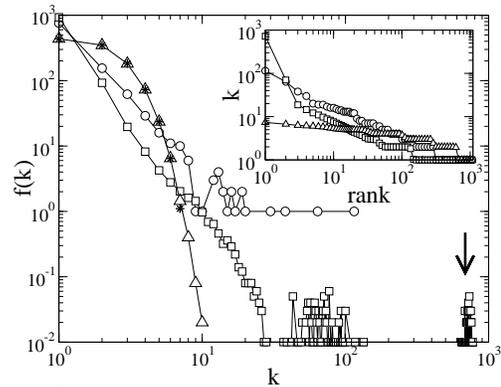}}
\caption{Frequency distribution of the degree of the MST of 
real data (circle). We also show the mean degree distribution
of random (triangle) and one-factor (square) model averaged 
over 100 numerical realizations of the MST.
The stars are the theoretical values of the degree frequency 
for the random model in mean field limit. 
The inset shows the corresponding rank plot of the degree in the three cases.}
\label{fig3}
\end{figure}

\begin{figure}
\centerline{\psfig{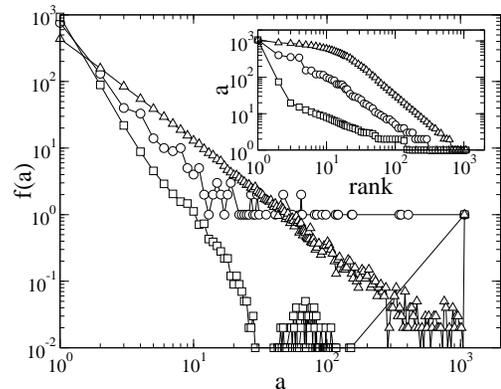}}
\caption{Frequency distribution of the in-degree component 
of the MST of real data (circle). We also show the mean 
in-degree component distribution 
of random (triangle) and one-factor (square) model
averaged over 100 numerical realizations of the MST.
The inset shows the corresponding rank plot of the in-degree component
for the three cases.}
\label{fig4}
\end{figure}

\end{document}